\documentclass[pra,floatfix,nofootinbib,reprint,superscriptaddress]{revtex4-1}

\usepackage{iopams,setstack,amsopn} 
\expandafter\let\csname equation*\endcsname\relax
\expandafter\let\csname endequation*\endcsname\relax
\usepackage{amsmath, amssymb, amsthm}
\usepackage{graphicx}
\usepackage{dcolumn}
\usepackage{times}
\usepackage{tabularx}
\usepackage{subfigure}
\usepackage[e]{esvect}
\usepackage{tikz}
\usetikzlibrary{matrix}
\usepackage[pdftex, colorlinks, allcolors=blue]{hyperref}

\DeclareSymbolFont{sfletters}{OML}{cmbrm}{m}{it}

\DeclareMathOperator{\Tr}{Tr}
\newcommand{\dket}[1]{|#1\rangle}
\newcommand{\dbra}[1]{\langle#1|}
\newcommand{\dbraket}[2]{\langle#1|#2\rangle}

\begin{document}

\title{Variational tight-binding method for simulating large superconducting circuits}

\author{D.~K.~Weiss}
\email{dkweiss@u.northwestern.edu}
\address{Department of Physics and Astronomy, 
Northwestern University, Evanston, Illinois 60208, USA}
\author{Wade DeGottardi}
\altaffiliation[Present Address: ]{Department of Physics and Astronomy, Texas Tech University, Lubbock, Texas 79409, USA}
\address{Northrop Grumman Corporation, Linthicum, Maryland 21090, USA}
\author{Jens Koch}
\address{Department of Physics and Astronomy, Northwestern University, Evanston, Illinois 60208, USA}
\author{D.~G.~Ferguson}
\address{Northrop Grumman Corporation, Linthicum, Maryland 21090, USA}

\date{\today}

\begin{abstract}
We generalize solid-state tight-binding techniques for the spectral analysis of large superconducting circuits. We find that tight-binding states can be better suited for approximating the low-energy excitations than charge-basis states, as illustrated for the interesting example of the current-mirror circuit. The use of tight binding can dramatically lower the Hilbert space dimension required for convergence to the true spectrum, and allows for the accurate simulation of larger circuits that are out of reach of charge basis diagonalization.
\end{abstract}

\maketitle

\section{Introduction}
Increasing coherence and noise resilience in superconducting qubits is a key requirement on the roadmap for developing the next generation of error-corrected quantum processors surpassing the NISQ era. Intrinsic noise protection in superconducting circuits has therefore become an important focus of research \cite{Doucot2002, Ioffe2002, Kitaev2006, Gladchenko2009, Doucot2012, Brooks2013, Bell2014, Groszkowski2018, DiPaolo2018, Gyenis2019, Smith2019, kalashnikov2020, Rymarz2021}. However, achieving simultaneous protection from depolarization and dephasing is impossible for small circuits like the transmon, and instead necessitates circuits with two or more degrees of freedom. Such larger circuits, especially of the size considered for the current-mirror circuit \cite{Kitaev2006} or rhombi lattice \cite{Gladchenko2009, Doucot2012}, pose significant challenges for the quantitative analysis of energy spectra and prediction of coherence times. Consequently, the development of more efficient numerical tools capable of solving for eigenstates and eigenenergies of large superconducting circuits has emerged as a vital imperative. Strategies recently introduced for that purpose include hierarchical diagonalization \cite{Kerman2020}, adaptive mode decoupling \cite{ding2020}, and DMRG methods \cite{Lee2003, Weiss2019, DiPaolo2019}. Here, we propose variational tight binding as another strategy complementing the former ones and illustrate its application.  

Since the Hilbert space dimension $d$ of even a single transmon circuit is infinite, it is not fully accurate to blame the ``growth" of $d$ for the challenges encountered with circuits of larger size. Nonetheless, when representing the Hamiltonian in a basis not specifically tailored for the problem at hand, the dimension of the \emph{truncated} Hilbert space typically grows exponentially when choosing the truncation level such that a particular level of convergence is reached. This turns the numerical diagonalization of the circuit Hamiltonian into a hard problem. An approach to address this challenge, which is also implicitly represented by the strategies mentioned above, consists of constructing basis states which more closely approximate the desired low-energy eigenstates from the very beginning. As long as construction of the tailored basis and decomposition of the Hamiltonian in that basis can be accomplished efficiently, this approach will allow for reduced truncation levels and hence enable coverage of circuit sizes otherwise inaccessible numerically. 

Our construction of such tailored basis states is based on the observation that low-lying eigenstates of superconducting circuits are often localized in the vicinity of minima of the potential energy, when expressed in terms of appropriate generalized-flux variables. If the potential energy is periodic or at least periodic along certain axes, then the situation resembles the setting of a particle in a periodic potential, as commonly encountered in solid-state physics when considering electrons inside a crystal lattice. In the regime where tunneling between atomic orbitals of different atoms is weak, tight-binding methods are appropriate for band structure calculations \cite{Slater, Lowdin1950}. An analogous treatment has previously been applied to small circuits; see, for example, the discussions of tunneling between minima in the flux qubit \cite{Orlando1999, Chirolli2006, Tiwari2007}, the derivation of an asymptotic expression for the charge dispersion in the transmon qubit \cite{Koch2007}, or the analysis of charge noise in the fluxonium circuit \cite{Mizel2020}. 
Chirolli and Burkard carry out a full tight-binding description of the low-energy physics of the flux qubit, considering Bloch sums of harmonic oscillator ground-state wavefunctions localized in each minimum at the half-flux sweet spot \cite{Chirolli2006}. Motivated by the new interest in circuits of increased size and complexity, we build upon this research in two specific ways. First, we consider multiple basis states in each minimum, to both improve ground-state energy estimates and extract excited-state energies. Second, we consider minima that are not necessarily identical, and introduce an efficient means of calculating matrix elements between states localized in such inequivalent minima. These techniques allow us to demonstrate that tight-binding methods can be adapted for efficient computation of energy spectra of large circuits.

Our paper is organized as follows. In Sec.~\ref{sec:theory} we review the  tight-binding treatment in solid-state physics and develop its adaptation for superconducting circuits. We then detail our numerical implementation, involving the calculation of tight-binding matrix elements using ladder operator algebra. In Sec.~\ref{sec:fluxqubit} we first apply our numerical method to the simple example of a flux qubit and illustrate that tight binding can accurately reproduce the well-known results for the spectrum. In Sec.~\ref{sec:currentmirror} we then apply tight binding to the current-mirror circuit with up to nine degrees of freedom, and demonstrate that tight binding outperforms the diagonalization of the Hamiltonian in charge-basis representation in terms of accuracy, convergence behavior, and memory efficiency. We close with our conclusions in Sec.~\ref{sec:conclusion}. 

\section{Tight Binding for Superconducting Circuits}
\label{sec:theory}

Research on superconducting qubits has repeatedly encountered physics familiar from models and phenomena in solid-state physics. Examples include the close connection between the Cooper pair box and a particle in a one-dimensional crystal, or the interpretation of the fluxonium Hamiltonian in terms of Bloch states subject to interband coupling \cite{Koch2009}. Another analogy, which points to the computational technique applied to circuits in this article, is the consideration of crystal electrons in the tight-binding limit. In this regime, tunneling between electronic orbitals of different atoms is weak, and linear combinations of atomic orbitals constructed by ``periodically repeating" localized wavefunctions serve as a meaningful basis. The tight-binding method then employs this basis in an approximate solution to the Schr\"odinger equation. An analogous scenario can be encountered for superconducting circuits as shown in Fig.~\ref{fig:compare_tb_ed_flux_qubit}. Minima of the potential energy may give rise to localized states that are only weakly connected by tunneling to partner states in other potential minima. The ``atomic orbitals" which we will refer to as ``local wavefunctions" in this case can be identified with the harmonic oscillator states associated with a local Taylor expansion around each minimum.
\begin{figure*}
    \centering
    \includegraphics[width=1.95\columnwidth]{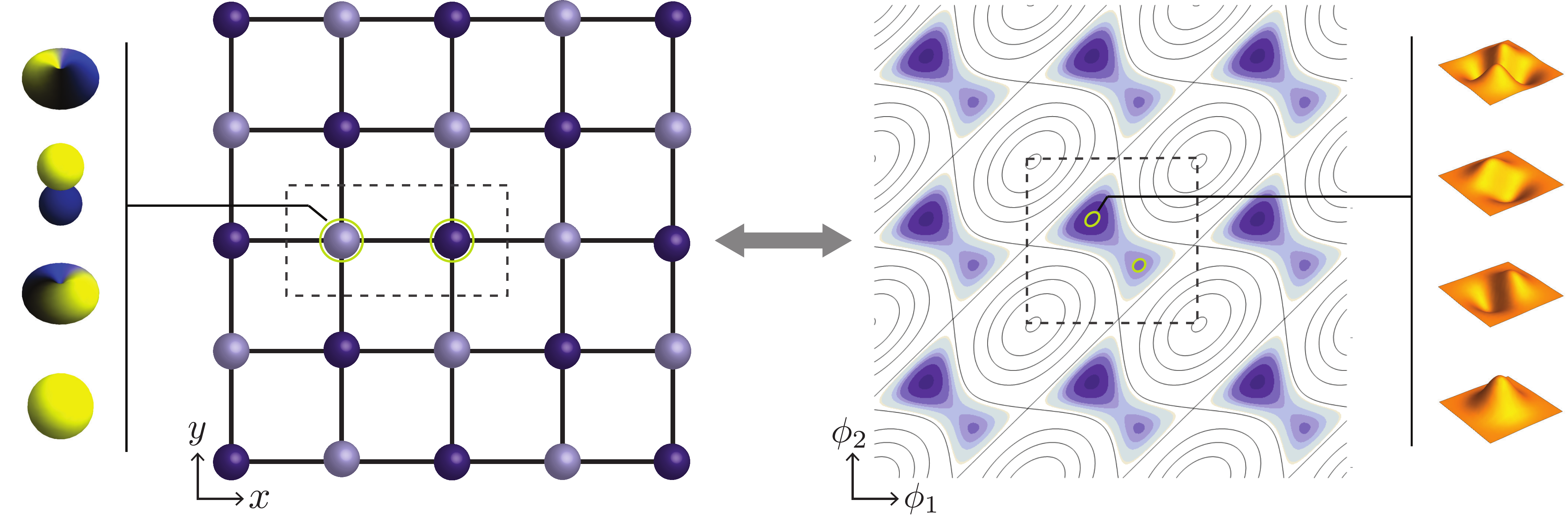}
    \caption{Comparison between tight binding as applied to solids and superconducting circuits. On the left is an example two-dimensional lattice with a two-atom basis, signified by the grey and blue atoms. On the far left are example atomic wavefunctions. On the right is the potential of the flux qubit, which has two inequivalent minima in each unit cell at the chosen value of flux. We only color the potential below a cutoff value to draw the eye to the potential minima locations. Near the minima the potential is approximately harmonic, therefore the local wavefunctions take the form of harmonic oscillator states. Example wavefunctions are shown on the right.}
    \label{fig:tb_compare}
\end{figure*}

\subsection{Local-wavefunction construction}
\label{sec:theory_A}

The starting point for this treatment is the full circuit Hamiltonian $\mathcal{H}=T+V$.
To stress the analogy with the setting of an infinite crystal, we first focus on a purely periodic potential $V(\vec{\phi})$, as realized by a circuit that does not include any inductors. (Including inductors is possible, which we comment on further in Sec.~\ref{sec:bloch}). In terms of the node variables $\vec{\phi}=(\phi_1, \ldots, \phi_{N})^{t}$, the potential energy obeys the periodicity condition $V(\vec{\phi}+2\pi\vec{j}) = V(\vec{\phi})$ with $\vec{j}\in \mathbb{Z}^N$ and thus forms a (hyper-)cubic Bravais lattice. Within the central unit cell defined by $\vec{\phi}\in [-\pi,\pi)^{\times N}$, the potential energy will exhibit a set of $M$ minima located at positions $\vec{\theta}_{m}$ where $m=0, 1, \ldots, M$. In the language of solid-state physics, this set of minima corresponds to the multi-atomic basis associated with the Bravais lattice. 

The analogy with solid-state physics is further strengthened by considering a gauge where the offset charge dependence is shifted from the Hamiltonian to the wavefunctions \cite{girvinleshouches, Chirolli2006}. In this representation, solutions $\dket{\psi}$ to the full Hamiltonian $\mathcal{H}$ obey quasiperiodic boundary conditions 
\begin{align}
\label{eq:quasiperiodic}
\mathcal{T}_{\vec{\theta}}\dket{\psi}=e^{-i\vec{n}_{g}\cdot\vec{\theta}}\dket{\psi},
\end{align}
for every $\vec{\theta}$ in the Bravais lattice, where $\mathcal{T}$ is the translation operator and $\vec{n}_{g}=(n_{g1}, \cdots, n_{gN})^{t}$ is the vector of offset charges. We recognize Eq.~\eqref{eq:quasiperiodic} as an expression of Bloch's theorem with wavevector $-\vec{n}_{g}$ (typically denoted as $\vec{k}$ in a solid-state context).

The construction of the local wavefunctions now proceeds by considering the individual harmonic oscillator Hamiltonians $\mathcal{H}_{m}=T+V_{m}$ where each local potential is obtained by Taylor expansion around the respective $m^{\text{th}}$ minimum,
\begin{align}
V_m=\frac{1}{2}\sum_{i,j}\varphi_{0}^2\phi_{i}^{(m)}\mathsf{\Gamma}^{(m)}_{ij}\phi_{j}^{(m)}.
\end{align}
Here, $\varphi_{0}= \hbar/2e$ is the reduced flux quantum, $\mathsf{\Gamma}_{ij}^{(m)}=\varphi_{0}^{-2}\partial_{\phi_{i}}\partial_{\phi_{j}}V|_{\vec{\theta}_{m}}$ the inverse of the inductance matrix and $\vec{\phi}^{(m)}=\vec{\phi}-\vec{\theta}_{m}$ the ``position" relative to the minimum location.
The local Hamiltonian then takes the form 
\begin{align}
\label{eq:local_ham}
\mathcal{H}_{m}'=\frac{1}{2}\sum_{i,j}\left(n_{i}8\frac{e^2}{2}(\mathsf{C})^{-1}_{ij}n_{j}
+\varphi_{0}^2\phi_{i}^{(m)}\mathsf{\Gamma}^{(m)}_{ij}\phi_{j}^{(m)}\right),
\end{align}
where $n_{i}$ is the charge number operator for node $i$ and $\mathsf{C}$ is the capacitance matrix. Hereafter, explicit references to $m$ will be omitted for notational simplicity. To obtain the eigenstates of the coupled oscillator Hamiltonian Eq.~\eqref{eq:local_ham} we first determine its normal modes. This is accomplished most efficiently based on the corresponding classical Lagrangian
\begin{align}
\label{eq:lagrangian}
\mathcal{L}'=\frac{1}{2}\varphi_{0}^2\sum_{i,j}\left(\dot{\phi}_{i}\mathsf{C}_{ij}\dot{\phi}_{j}-\phi_{i}\mathsf{\Gamma}_{ij}\phi_{j}\right).
\end{align}
Using the usual oscillatory solution ansatz $\vec{\phi}=\vec{\xi}_{\mu}e^{-i\omega_{\mu}t}$ reduces the equations of motion to the generalized eigenvalue problem $\mathsf{\Gamma}\vec{\xi}_{\mu}=\omega_{\mu}^2\mathsf{C}\,\vec{\xi}_{\mu}$ \cite{Goldstein}. Here, Latin indices refer to node variables, and Greek indices to normal mode variables. The eigenmode vectors $\vec{\xi}_{\mu}$ are only determined up to normalization, $\vec{\xi}_{\mu}^{\;T} \mathsf{C}\,\vec{\xi}_{\mu} = c_{\mu}$, implying $\vec{\xi}_{\mu}^{\;T} \mathsf{\Gamma}\vec{\xi}_{\mu} = \omega_{\mu}^2c_{\mu}$, where $c_{\mu}$ is undetermined. This normalization will be fixed when we return to the quantum-mechanical description, in such a way that the Hamiltonian for each mode takes the standard form 
\begin{align}
\label{eq:single_ham}
\mathcal{H}_{\mu}'/\hbar\omega_{\mu}=\frac{1}{2}\left(-\frac{\partial^2}{\partial\zeta_{\mu}^2}+\zeta_{\mu}^2 \right).
\end{align}
Here, $\vec{\zeta}=(\zeta_{\mu})$ collects the normal-mode variables related to the original generalized fluxes via $\vec{\phi}=\Xi\vec{\zeta}$, where $\Xi$ is the matrix of column vectors $\vec{\xi}_{\mu}$.\footnote{Note that the matrix $\Xi$ encodes both the normal-mode directions and oscillator lengths. For the 1D example of a transmon within the harmonic approximation, $\mathcal{H}'_{\text{tran}}=4E_{C}n^2+\frac{1}{2}E_{J}\phi^{2}$, the matrix $\Xi$ reduces to the number $(8E_{C}/E_{J})^{1/4}$ which indeed is the corresponding harmonic length.} In these new variables, both bilinear forms in $\mathcal{L}'$ are diagonal.
Legendre transform and quantization thus readily yield
\begin{align}
\label{eq:local_ham_diag}
\mathcal{H}'=\frac{1}{2}\sum_{\mu}\left[-\left(\frac{\hbar}{\varphi_{0}}\right)^2\frac{1}{c_{\mu}}\frac{\partial^2}{\partial\zeta_{\mu}^2}+\left(\varphi_{0}\omega_{\mu}\right)^2c_{\mu}\zeta_{\mu}^2 \right].
\end{align}
To cast $\mathcal{H}'$ into the form suggested by Eq.~\eqref{eq:single_ham} we now choose $c_{\mu}=(2e)^2/\hbar\omega_{\mu}$ as our normalization constants.
We denote the eigenstates of $\mathcal{H}'$ by $\dket{\vec{s},m}$. Here, $\vec{s}$ collects the excitation numbers $s_{\mu}=0,1,\ldots$ for each mode $\mu$ and $m$ specifies the minimum of interest.

\subsection{Bloch summation and the generalized eigenvalue problem}
\label{sec:bloch}

Solution of the Schr\"odinger equation $\mathcal{H}\dket{\psi}=E\dket{\psi}$ proceeds by choosing a basis with which to express $\mathcal{H}$ in matrix form. We construct this basis by periodic repetition over the entire Bravais lattice of the local wavefunctions $\dket{\vec{s},m}$ defined in the central unit cell, subject to quasiperiodic boundary conditions [Eq.~\eqref{eq:quasiperiodic}]
\begin{align}
\label{eq:ket_quasiperiodic}
\dket{\psi_{\vec{n}_{g},\vec{s},m}}&=\frac{1}{\sqrt{\mathcal{N}}}\sum_{\vec{j}}e^{-i\vec{n}_{g}\cdot(2\pi\vec{j}+\vec{\theta}_{m})}\mathcal{T}_{2\pi\vec{j}}\dket{\vec{s},m}, \\ \nonumber
&=\frac{1}{\sqrt{\mathcal{N}}}\sum_{\vec{j}}\dket{\vec{s},m;\vec{j}}.
\end{align}
Here, $\mathcal{N}$ is the number of unit cells and $\dket{\vec{s},m;\vec{j}}$ the wavefunction localized in minimum $m$ in the unit cell located at $2\pi\vec{j}$. (Note that these kets $\dket{\vec{s},m;\vec{j}}$ are implictly offset-charge dependent). 
It is straightforward to show that $\dket{\psi_{\vec{n}_{g},\vec{s},m}}$ satisfies the quasiperiodicity condition \eqref{eq:quasiperiodic}. We now represent the Schr\"odinger equation in terms of these basis states. Due to their lack of orthogonality, this transforms the Schr\"odinger equation into the generalized eigenvalue problem
\begin{align}
\label{eq:first_gen}
\frac{1}{\mathcal{N}}\sum_{\vec{s},m}\sum_{\vec{j},\vec{j}\,'}\Big( &\dbra{\vec{s}\,',m';\vec{j}\,'}\mathcal{H}\dket{\vec{s}, m;\vec{j}\,} \\ \nonumber -& E \dbraket{\vec{s}\,',m';\vec{j}\,'}{ \vec{s},m;\vec{j}\,}\Big)b_{\vec{s},m}=0,
\end{align}
where $E$ is the eigenenergy and $b_{\vec{s},m}$ are the coefficients in the decomposition
\begin{align}
\label{eq:tightbindingansatz}
\dket{\psi_{\vec{n}_{g}}}= \sum_{\vec{s},m}b_{\vec{s},m}\dket{\psi_{\vec{n}_{g},\vec{s},m}}.
\end{align}
Eq.~\eqref{eq:first_gen} can be simplified by performing one of the sums over lattice vectors \cite{Slater}. This can be done by expressing the kets explicitly in terms of the translation operators $\dket{\vec{s},m;\vec{j}}=e^{-i\vec{n}_{g}\cdot(2\pi\vec{j}+\vec{\theta}_{m})}\mathcal{T}_{2\pi\vec{j}}\dket{\vec{s},m;\vec{0}}$ and noting that the operator $\mathcal{T}_{2\pi\vec{j}}$ commutes with the Hamiltonian. The summation yields a factor of $\mathcal{N}$ and we obtain
\begin{align}
\label{eq:tightbindingHamiltonian}
\sum_{\vec{s},m}\sum_{\vec{j}}\Big( &\dbra{\vec{s}\,',m';\vec{0}\,}\mathcal{H}\dket{\vec{s}, m;\vec{j}\,} \\ \nonumber -& E \dbraket{\vec{s}\,',m';\vec{0}\,}{ \vec{s},m;\vec{j}\,}\Big)b_{\vec{s},m}=0.
\end{align}
Formally, Eq.~\eqref{eq:tightbindingHamiltonian} now has the standard form of a generalized eigenvalue problem with two semidefinite positive Hermitean matrices and can be handled numerically by an appropriate solver. To accomplish this, the crucial remaining task consists of the efficient evaluation of the matrix elements and state overlaps in Eq.~\eqref{eq:tightbindingHamiltonian}.
Note that an alternative route to this equation is application of the variational principle to $\dbra{\psi_{\vec{n}_{g}}}\mathcal{H}\dket{\psi_{\vec{n}_{g}}}=E\dbraket{\psi_{\vec{n}_{g}}}{\psi_{\vec{n}_{g}}}$ \cite{Bishop1989}; the benefit of this viewpoint is that the eigenenergies thus obtained represent upper bounds to the true eigenenergies of the system \cite{Bishop1989, Macdonald1933}.

Our analysis thus far has assumed a purely periodic potential, allowing for a direct analogy with the theory of tight binding as applied to solids. Including inductive terms in the potential immediately implies that associated degrees of freedom are no longer subject to (quasi-)periodic boundary conditions. Alternatively, we can say that the unit cell no longer has finite volume, but must extend along the relevant axes. To include such inductive potential terms, we therefore do not perform periodic summation in Eq.~\eqref{eq:ket_quasiperiodic} along these non-periodic directions. We have successfully implemented the tight-binding method for the symmetric $0-\pi$ qubit \cite{Brooks2013, Groszkowski2018}, a circuit with one periodic and one extended degree of freedom. The  low-energy spectra thus obtained are in excellent agreement with exact results over a wide range of circuit parameters. In this paper we will continue to focus on circuits with purely periodic potentials, the natural setting for the tight-binding method. We defer a detailed discussion of our results for the $0-\pi$ qubit to a future publication.

\subsection{Efficient computation of matrix elements and overlaps}

The relevant matrix elements involve harmonic-oscillator states at different locations and, possibly, with different normal-mode orientations and oscillator lengths. The calculation of these quantities proceeds either via use of ladder operators or by explicit integration within the position representation. Even though integration can in principle be accomplished analytically, the expressions become increasingly tedious in higher dimensions. (The integrals are generally two-center integrals that lead to two-variable Hermite polynomials \cite{Babusci2012}). By contrast, the ladder-operator formalism is more readily adapted for the numerical calculations of the matrix elements in question. Therefore, we focus on this approach. 

The matrix elements and overlaps to be evaluated have the form
\begin{align}
\label{eq:matelem}
\dbra{\vec{s}\,',m';\vec{0}\,}\mathcal{O}\dket{\vec{s},m;\vec{j}\,},
\end{align}
where $\mathcal{O}$ is either the Hamiltonian $\mathcal{H}$ or the identity. To facilitate the use of the ladder-operator formalism, we next re-express operators and states in terms of the creation and annihilation operators associated with the $m=0$ minimum in the central unit cell.
Since inequivalent minima differ in locations and curvatures, local wavefunctions are shifted and possibly squeezed relative to each other,
\begin{align}
\label{eq:squeezing}
\mathcal{T}_{\vec{\theta}_{m}}\mathcal{S}_{m}\dket{\vec{s}\,}=\dket{\vec{s}\,,m},
\end{align}
where $\dket{\vec{s}\,}\equiv\dket{\vec{s},0}$ and we have taken the location of the $m=0$ minimum to be the origin, $\vec{\theta}_{m=0}=\vec{0}$. The intuitive interpretation of Eq.~\eqref{eq:squeezing} is based on a two-step process: first the harmonic oscillator states for $m=0$ are deformed to match the local curvature of the $m^{\text{th}}$ minimum and are then shifted over to the appropriate location of that minimum.
According to Eq.~\eqref{eq:squeezing}, the matrix elements take the form
\begin{align}
\label{eq:matelem_2}
\dbra{\vec{s}\,',m';\vec{0}\,}\mathcal{O}\dket{\vec{s},m;\vec{j}\,} = \dbra{\vec{s\,}'}\mathcal{S}_{m'}^{\dagger}\mathcal{T}_{\vec{\theta}_{m'}}^{\dagger}\mathcal{O}\mathcal{T}_{\vec{\theta}_{m}+2\pi\vec{j}}\mathcal{S}_{m}\dket{\vec{s}\,}.
\end{align}
The expression for the states is readily obtained,
\begin{align}
\label{eq:state_ladder}
\dket{\vec{s}\,} = \prod_{\mu}\frac{1}{\sqrt{s_{\mu}!}}(a_{\mu}^{\dagger})^{s_{\mu}}\dket{\vec{s}=\vec{0}\,}.
\end{align}
Likewise the treatment of the translation operator $\mathcal{T}_{\vec{\theta}}=e^{-i\vec{\theta}\cdot\vec{n}}$ is straightforward: making use of the relation between the number operators and ladder operators 
\begin{align}
\label{eq:naad}
n_{j}=\sum_{\mu}\frac{-i}{\sqrt{2}}\Xi_{j\mu}^{-t}(a_{\mu}-a_{\mu}^{\,\dagger}),
\end{align}
the translation operator can be expressed as 
\begin{align}
\label{eq:T}
\mathcal{T}_{\vec{\theta}}
= \exp(-\tfrac{1}{\sqrt{2}}\vec{\theta}^{\,t}\,\Xi^{-t}[\vec{a}-\vec{a^{\dagger}}]).
\end{align}
Here we use the compact notation $\vec{a}=(a_{1},\cdots,a_{N})^{t}$ and $\vec{a^{\dagger}}=(a_{1}\,^{\dagger},\cdots,a_{N}\,^{\dagger})^{t}$ and denote $\Xi^{-t}=(\Xi^{-1})^{t}$.

The expression for the squeezing operator can be found by considering a simplified situation of two harmonic Hamiltonians $\mathcal{H}_{a}, \mathcal{H}_{c}$ of the form of Eq.~\eqref{eq:local_ham}, but defined at the same center point. The Hamiltonian $\mathcal{H}_{a}$ is diagonalized by the ladder operators $\vec{a}, \vec{a^{\dagger}}$ and $\mathcal{H}_{c}$ is diagonalized by $\vec{c}, \vec{c^{\dagger}}$. The respective eigenfunctions $\dket{\vec{s}\,}_{a}$ and $\dket{\vec{s}\,}_{c}$ are related by a unitary squeezing transformation,
\begin{align}
\label{eq:squeezing_definition}
\mathcal{S}\dket{\vec{s}\,}_{a}=\dket{\vec{s}\,}_{c},
\end{align}
which is equivalent to 
$\mathcal{S}\vec{a}\mathcal{S}^{\dagger} = \vec{c}$.
To obtain a concrete expression for $\mathcal{S}$ we note that
the bosonic ladder operators are related by a Bogoliubov transformation \cite{Javanainen, Bogoliubov, *Valatin}
\begin{align}
\label{eq:bogoliubov}
\left(\begin{matrix}u & v \\ v^* & u^* \end{matrix} \right)
\left(\begin{matrix}\vec{a} \\ \vec{a^{\dagger}} \end{matrix}\right)=\left(\begin{matrix}\vec{c} \\ \vec{c^{\dagger}} \end{matrix}\right),
\end{align}
where $u, v$ are $N\times N$ matrices.
These can be found by considering the two decompositions of the phase and number operators in terms of the differing sets of ladder operators
\begin{align}
\label{eq:phi_n_a}
\left(\begin{matrix}\vec{\phi} \\ \vec{n} \end{matrix}\right) &= \frac{1}{\sqrt{2}}
\left(\begin{matrix} \Xi & \Xi \\ -i\Xi^{-t} & i\Xi^{-t} \end{matrix} \right)
\left(\begin{matrix}\vec{a} \\ \vec{a^{\dagger}} \end{matrix}\right)
\\ \nonumber
&=\frac{1}{\sqrt{2}}\left(\begin{matrix} \Xi' & \Xi' \\ -i\Xi'^{-t} & i\Xi'^{-t} \end{matrix} \right)\left(\begin{matrix}\vec{c} \\ \vec{c^{\dagger}} \end{matrix}\right),
\end{align}
where the matrix $\Xi$ is defined for $\mathcal{H}_{a}$ and $\Xi'$ for $\mathcal{H}_{c}$ as in Sec.~\ref{sec:theory_A}.
Solving Eq.~\eqref{eq:phi_n_a} for the ladder operators $\vec{c}, \vec{c^{\dagger}}$ yields the real-valued Bogoliubov matrices
\begin{align}
\substack{\displaystyle u\\[1mm]\displaystyle v}
&= \frac{1}{2}(\Xi'^{-1}\Xi \pm \Xi'^{t}\Xi^{-t}).
\end{align}
As shown in Ref.~\cite{Balian1969}, the multimode squeezing operator can now be expressed in terms of $u, v$ as follows:
\begin{align}
\label{eq:squeezingop}
\mathcal{S} = \exp(\tfrac{1}{2}(\vec{a}\,^{t}\;\vec{a^{\dagger}}^{t})J\ln{M}(\vec{a}\;\vec{a^{\dagger}})^{t}),
\end{align}
where
\begin{align}
M = \left(\begin{matrix}u & v \\ v & u \end{matrix} \right), \;\;J=\left(\begin{matrix}\phantom{-}0 & \openone \\ -\openone & 0 \end{matrix} \right).
\end{align}
Returning now to our original notation, we identify $\mathcal{S}_{m}$ with $\mathcal{S}$ in Eq.~\eqref{eq:squeezingop}, where the $m$ dependence carries forward to the Bogoliubov $u, v$ matrices. With Eqs.~\eqref{eq:state_ladder},\eqref{eq:T},\eqref{eq:squeezingop} we have, in principle, collected all ingredients necessary for the evaluation of the matrix elements and overlaps [Eq.~\eqref{eq:matelem_2}]. However, numerical implementation necessarily involves truncation, and we will show in the following that normal-ordering operator expressions is essential for maximizing accuracy.

A standard approach for truncating the infinite-dimensional operators $a_{\mu}, a_{\mu}^{\dagger}$ consists of excitation cutoffs $s_{\text{max}}$ applied to each mode. This is a fine strategy for moderate sized systems, but quickly becomes intractable for larger systems. To mitigate this bottleneck, one can instead use a global excitation number cutoff $\Sigma_{\text{max}}$, which institutes a maximum Manhattan length of the excitation number vector, $\|\vec{s}\,\|_{1}\leq\Sigma_{\text{max}}$ \cite{Zhang_2010}.

Given a specific truncation level, it makes a difference whether operator expressions are normal ordered or not.
Denoting the truncated operators as $\tilde{a}_{\mu}, \tilde{a}_{\mu}^{\dagger}$, the nominally identical expressions $\tilde{a}_{\mu}\tilde{a}_{\mu}^{\dagger}$ and $\tilde{a}_{\mu}^{\dagger}\tilde{a}_{\mu}+\delta_{\mu\mu}$ in fact give different results as seen, for instance, in
\begin{align}
\dbra{s_{\text{max}}}\tilde{a}_{\mu}\tilde{a}_{\mu}^{\dagger}\dket{s_{\text{max}}} &=0, \\ \nonumber 
\dbra{s_{\text{max}}}(\tilde{a}_{\mu}^{\dagger}\tilde{a}_{\mu}+\delta_{\mu\mu})\dket{s_{\text{max}}} &=s_{\text{max}}+1.
\end{align}
Here, the ``wrong" result of the first expression can be circumvented by using the normal-ordered version in the second expression. This example is indicative of a general result, that it is beneficial to normal order ladder-operator expressions before further numerical evaluation.

The translation operator $\mathcal{T}_{\vec{\theta}}$ can be normal ordered via the the Baker-Campbell-Hausdorff (BCH) formula \cite{Baker, *Campbellone, *Campbelltwo, *Hausdorff}, which takes the form 
$e^{X}e^{Y}=e^{X+Y+\frac{1}{2}[X,Y]}$ when $X$ and $Y$ are operators that commute with their commutator $[X, Y]$. This yields
\begin{align}
\label{eq:BCH}
\mathcal{T}_{\vec{\theta}}
= &\mathcal{V}_{\vec{\theta}}^{\dagger}\,\mathcal{V}_{-\vec{\theta}}\,\exp(-\tfrac{1}{4}\vec{\theta}^{\,t}\Xi^{-t}\Xi^{-1}\vec{\theta}),
\end{align}
where
\begin{align}
\label{eq:V}
\mathcal{V}_{\vec{\theta}} = \exp(\tfrac{1}{\sqrt{2}}\vec{\theta}^{\,t}\Xi^{-t}\vec{a}).
\end{align}
Expressions for commuting $\mathcal{V}$ operators past operators such as $n_{j}$ and $e^{i\phi_{j}}$ (which enter $\mathcal{O}$ in Eq.~\eqref{eq:matelem}) can be easily obtained:
\begin{align}
\label{eq:normalkinetic}
\mathcal{V}_{\vec{\theta}}\,\,\vec{n} &=\left(\vec{n}+\frac{i}{2}\Xi^{-t}\Xi^{-1}\vec{\theta}\right)\mathcal{V}_{\vec{\theta}}, \\ 
\label{eq:normalpotential}
\mathcal{V}_{\vec{\theta}}\,e^{i\phi_{j}} &=e^{i(\phi_{j}+\frac{1}{2}\theta_{j})}\mathcal{V}_{\vec{\theta}},
\end{align}
where in Eq.~\eqref{eq:normalkinetic} we have used the identity \cite{Wilcox} $e^{X}Y = (Y+[X,Y])e^{X}$, again valid when $X$ and $Y$ commute with $[X, Y]$.

The normal-ordering procedure for the squeezing operators is more involved and we defer it to Appendix ~\ref{appendix:squeezing}. In the following sections where we apply the tight-binding method to several example systems, we find that the sets of basis states constructed with (proper tight binding) or without (improper tight binding) squeezing may yield similar numerical performance. This is naturally the case if minima contributing to the low-energy spectrum have similar curvatures. Whenever possible, omitting squeezing from the construction of basis states significantly simplifies the numerical treatment.

The final step in setting up the generalized eigenvalue problem \eqref{eq:tightbindingHamiltonian}, is truncating the sum over vectors $\vec{j}$. A typical truncation scheme is the nearest-neighbor approximation which selects only those unit cells that have the minimal Euclidean distance from the central unit cell. This strategy however does not account for any anisotropy in the harmonic lengths, which results in local wavefunctions whose Gaussian tails extend further in some directions than in others. We therefore use a different criterion based on the overlap of local wavefunctions. Whether the unit cell centered at $2\pi\vec{j}$ is a nearest neighbor to the central unit cell now generally depends on the minima under consideration. Specifically, given a minimum $m'$ in the central unit cell, and a minimum $m$ in the unit cell at vector  $2\pi\vec{j}$, we determine the nearest-neighbor character by computing the overlap of the two harmonic oscillator ground-state wavefunctions. For a given overlap threshold value $\epsilon$, we call the two unit cells nearest neighbors with respect to $m$ and $m'$ if
\begin{align}
\label{eq:nearestneighbor}
\dbraket{\vec{0},m';\vec{0}}{\vec{0},m;\vec{j}\,} =&
\sqrt{\frac{2^N\det(\Delta_{m})^{1/2}\det(\Delta_{m'})^{1/2}}{\det(\Delta_{m}+\Delta_{m'})}} \\ \nonumber 
\times& \exp(-\tfrac{1}{2}\vec{\delta\theta}^{t}(\Delta_{m}^{-1} + \Delta_{m'}^{-1})^{-1}\vec{\delta\theta})
\\ \nonumber >& \;\epsilon.
\end{align}
Here, we have defined $\vec{\delta\theta}=2\pi\vec{j} + \vec{\theta}_{m} - \vec{\theta}_{m'}$ and $\Delta_{m}=\Xi_{m}^{-t}\Xi_{m}^{-1}$, where $\Xi_{m}$ is defined relative to minimum $m$. 
With this definition in place, we truncate the sum over $\vec{j}$ by selecting neighbors up to a certain degree. (Note that the overlap threshold $\epsilon$ must ultimately be adjusted adaptively in order to ensure convergence.)

A possible challenge for the numerical treatment, which we have observed in several cases, is that the overlap matrix $\langle\psi_{\vec{n}_{g},\vec{s}\,',m'}|\psi_{\vec{n}_{g},\vec{s},m}\rangle$ may approach singularity (and possibly become indefinite due to rounding errors). This is a familiar problem in quantum chemistry calculations \cite{reed1988, lu2012} and arises when the set of ``basis" states $\{\dket{\psi_{1}},\dket{\psi_{2}},\ldots ,\dket{\psi_{h}}\}$ is approximately linearly dependent. A common technique for resolving this issue which we have implemented here is the {\it canonical orthogonalization} procedure of L\"owdin  \cite{Lowdin1956, *Lowdin1967, *Lowdin1970}. One diagonalizes the inner product matrix to obtain the eigenvalues $\{\Delta_1, \Delta_2,\ldots,\Delta_h \}$ and matrix of column eigenvectors $U$. The orthonormalized states are \cite{Lowdin1956, *Lowdin1967, *Lowdin1970}
\begin{align}
\dket{\psi'_{k}}=\Delta_{k}^{-1/2}\sum_{\ell}\dket{\psi_{\ell}}U_{\ell k}.
\end{align}
Choosing a cutoff $\Delta_{\text{min}}$ allows for the rejection of states $\dket{\psi'_{k}}$ where $\Delta_{k} < \Delta_{\text{min}}$. The Hamiltonian $\mathcal{H}$ is then projected onto the deflated basis and we are left with a standard eigenvalue problem.

\subsection{Optimization and anharmonicity correction of the ansatz wavefunctions}
\label{sec:anharmonic_correction}

\begin{figure}
    \centering
    \includegraphics[width=\columnwidth]{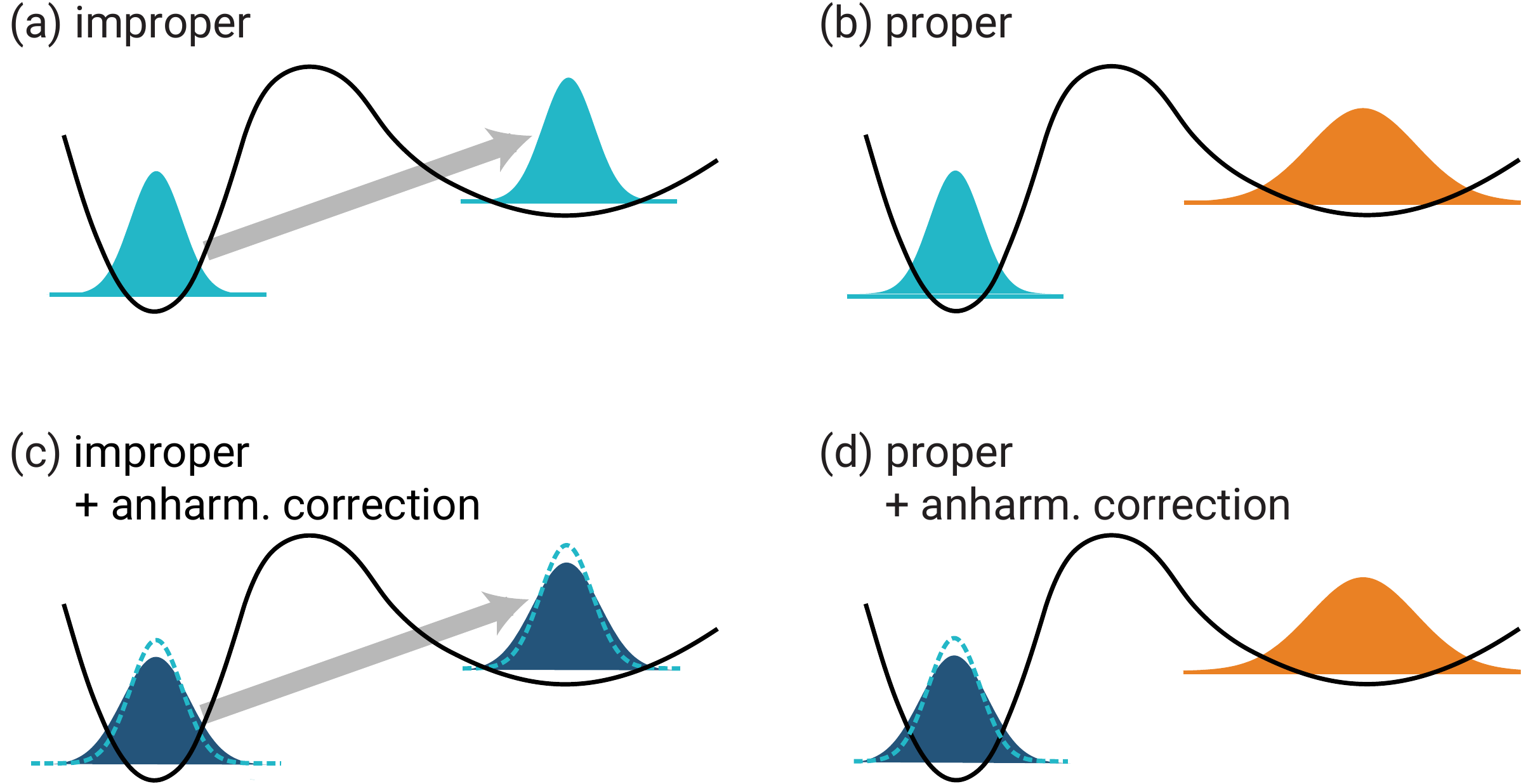}
    \caption{Schematic of ansatz construction schemes. (a) Improper, where local wavefunctions are defined according to the curvature of the $m=0$ minimum and are reused to form the local wavefunctions of other inequivalent minima. (b) Proper, where local wavefunctions for every minimum are defined according to the local curvature. (c) Improper with anharmonicity correction, where harmonic length(s) of the ansatz ground-state wavefunction of the $m=0$ minimum are optimized to account for anharmonicity corrections to the potential. The resulting wavefunctions are then also used for $m\neq0$ minima as in (a). The dashed lines show the unoptimized local ground state wavefunction defined for the $m=0$ minimum (the change in the harmonic length due to anharmonicity corrections has been exaggerated). (d) Proper with anharmonicity correction of only the $m=0$ minimum. Wavefunctions for the $m=0$ minimum are defined according to the local curvature and anharmonicity correction scheme, while wavefunctions for $m\neq0$ minima are defined only according to the local curvature. }
    \label{fig:TB_schemes}
\end{figure}

One of the main goals of this work is the construction of basis states that closely approximate the low-energy eigenstates of superconducting circuits. We can
optimize the tight-binding wavefunctions \eqref{eq:ket_quasiperiodic} for this purpose by recognizing that sufficiently far from each minimum location, the potential ceases to be strictly harmonic. The low-energy eigenfunctions typically have spatial spreads that are broader if the leading-order anharmonic term is negative and narrower if it is positive. We take this effect into account and improve the tight-binding wavefunctions by treating the harmonic length of each mode as a variational parameter. Specifically, we modify the matrix $\Xi$ by optimizing the magnitude of the eigenvectors $\vec{\xi}_{\mu}$, leaving the directions unchanged, $\vec{\xi}_{\mu}\rightarrow\lambda_{\mu}\vec{\xi}_{\mu}$, where $\lambda_{\mu}$ is optimized. We perform this optimization procedure for the ansatz ground state $\dket{\psi_{\vec{n}_{g}, \vec{0}, m}(\lambda_{\mu})}$, making the dependence on $\lambda_{\mu}$ explicit, minimizing
\begin{align}
E=\frac{\dbra{\psi_{\vec{n}_{g}, \vec{0}, m}(\lambda_{\mu})}\mathcal{H}\dket{\psi_{\vec{n}_{g}, \vec{0}, m}(\lambda_{\mu})}}{\dbraket{\psi_{\vec{n}_{g}, \vec{0}, m}(\lambda_{\mu})}{\psi_{\vec{n}_{g}, \vec{0}, m}(\lambda_{\mu})}}.
\end{align}
The resulting harmonic lengths are then used for all other states defined in the same minimum $m$.\footnote{Alternatively, one could optimize the harmonic lengths of a higher-lying basis state \cite{Bishop1989}, which is a possible avenue for future research.} 
We term this optimization scheme ``anharmonicity correction,"
which combined with improper and proper tight binding leads to additional choices for constructing tight-binding states: (IPAC) {\bf i}m{\bf p}roper with {\bf a}nharmonic {\bf c}orrection and (PAC) {\bf p}roper with {\bf a}nharmonic {\bf c}orrection of the $m=0$ minimum, see Fig.~\ref{fig:TB_schemes}. 
We could further envision the construction of states according to: proper with anharmonicity correction of all minima. However, we find that this scheme yields no numerical benefit over the other methods in all cases considered here, thus we do not discuss it further.

\subsection{Applicability of tight binding}
\label{sec:applicability}

A natural question to ask is whether the tight-binding method is appropriate for obtaining the eigenspectrum of a given superconducting circuit. A general and systematic answer to this question is difficult to obtain and we do not aim to give a comprehensive answer here. Instead we seek to motivate a ``rule-of-thumb" criteria that serves as an indicator of whether the tight-binding method can produce meaningful results.

If the spatial spread of the localized harmonic-oscillator states [eigenfunctions of Eq.~\eqref{eq:local_ham_diag}] is small compared to distances between minima then the tight-binding approach is physically well motivated and we expect tight-binding wavefunctions to serve as good approximations to low-energy eigenstates. If, on the other hand, the wavefunctions have large spatial spread and significant overlap, then the weak-periodic-potential approximation is more appropriate for describing the low-energy excitations.

To quantify this discussion, we define length scales to compare the spatial spread of wavefunctions with the distance between minima. Examining the exponential dependence $\exp(-\frac{1}{2}\vec{\phi}^{\,t}\Xi^{-t}\Xi^{-1}\vec{\phi})$ of the local harmonic wavefunctions, we can extract the effective harmonic length $\ell_{mm'}$ along the unit vector $\hat{u}_{mm'}$ separating two minima $m$ and $m'$
\begin{align}
\ell_{mm'}\equiv(\hat{u}_{mm'}^{t}\Xi^{-t}\Xi^{-1}\hat{u}_{mm'})^{-1/2}.
\end{align}
It is natural to compare $\ell_{mm'}$ to $d_{mm'}/2$, half the distance between the minima. Our rule-of-thumb for application of the tight-binding method is based on the smallness of the localization ratios
\begin{align}
r_{mm'}=\frac{d_{mm'}/2}{\ell_{mm'}},
\end{align}
compared to unity. This provides a rough threshold for judging whether the tight-binding method might be appropriate.

\section{Tight binding applied to the flux qubit}
\label{sec:fluxqubit}
In order to evaluate the accuracy of the tight-binding method, we first apply it to the familiar case of the three-junction flux qubit. The spectrum of the flux qubit is well understood \cite{Orlando1999, You2007}, but applying the method in this context is of interest and nontrivial because the flux qubit has multiple degrees of freedom and multiple inequivalent minima in the central unit cell. Additionally, the flux qubit is typically operated in a parameter regime where tight-binding techniques are applicable. Indeed, many authors have used tight-binding techniques to get analytical estimates of tunneling rates and low-energy eigenvalues \cite{Orlando1999, Chirolli2006, Tiwari2007}. We extend this previous research by using multiple tight-binding basis states in each inequivalent minimum to obtain improved low-energy eigenvalue estimates.

We consider the case where two of the junctions are identical with junction energy $E_{J}$ and capacitance $C_{J}$, while the third has junction energy and capacitance reduced by a factor of $\alpha$. The Hamiltonian is \cite{Orlando1999}
\begin{align}
\label{eq:fluxqubithamiltonian}
\mathcal{H}_{\text{flux}}=& \sum_{i,j=1}^{2}(n_{i}-n_{gi})4(\mathsf{E_{C}})_{ij}(n_{j}-n_{gj})\\ \nonumber -& E_{J}\cos(\phi_{1})-E_{J}\cos(\phi_{2}) \\ \nonumber  
-& \alpha E_{J}\cos(\phi_{1}-\phi_{2} + \varphi_{\text{ext}}) + E_{J}(2+\alpha),
\end{align}
where $\varphi_{\text{ext}}=2\pi\Phi_{\text{ext}}/\Phi_{0}$, $\Phi_{0}=h/2e$ is the flux quantum, $\mathsf{E_{C}}=\frac{e^2}{2}\mathsf{C}^{-1}$ is the charging energy matrix and the constant term is included to ensure that the spectrum of $H_{\text{flux}}$ is positive. 
The capacitance matrix $\mathsf{C}$ is
\begin{align}
\mathsf{C}=\left(\begin{matrix}
C_{J}(1+\alpha)+C_{g} & -\alpha C_{J} \\
-\alpha C_{J} & C_{J}(1+\alpha)+C_{g}
\end{matrix}\right),
\end{align}
where $C_{g}$ is the capacitance to ground of each island. (See Refs.~\cite{Orlando1999, You2007} for details on the derivation of Eq.~\eqref{eq:fluxqubithamiltonian}).

\begin{figure}
    \centering
    \includegraphics[width=1.0\columnwidth]{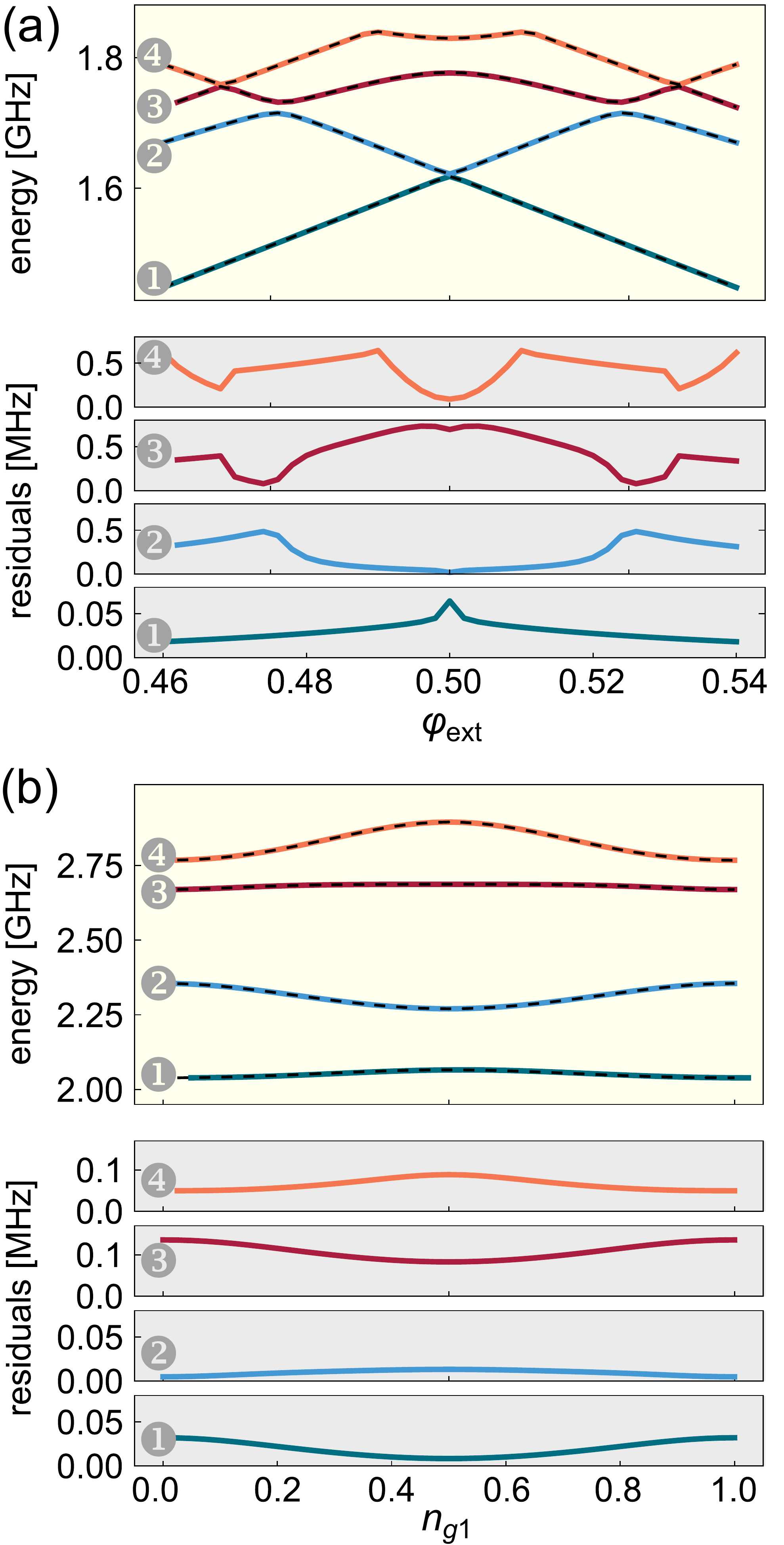}
    \caption{Spectrum of the flux qubit as a function of (a) flux and (b) offset charge $n_{g1}$, calculated using charge-basis diagonalization (solid) and improper tight binding (dashed). At the magnification level of the two figures, the spectra almost exactly overlap. Below each spectrum is the absolute error of the tight-binding calculation relative to the exact spectrum for each of the four lowest-energy eigenstates. Sub-MHz level agreement is achieved in all cases considered here with $\Sigma_{\text{max}}=5$, and sub-kHz level absolute error is possible for both parameter sets by increasing $\Sigma_{\text{max}}$.
    For (a) flux modulation, we choose parameters $E_{J}/h=1$ GHz, $E_{J}/E_{C_{J}}=60, E_{C_{g}}/E_{C_{J}}=50, \alpha=0.8$ and $n_{gi}=0$ \cite{You2007}. For (b) tuning $n_{g1}$, we use parameters $E_{J}/h=1$ GHz, $E_{J}/E_{C_{J}}=5, E_{C_{g}}/E_{C_{J}}=50, \alpha=0.8$ and have set $n_{g2}=0$, $\varphi_{\text{ext}}=0.5$.}
    \label{fig:compare_tb_ed_flux_qubit}
\end{figure}

In order to demonstrate quantitative accuracy of the tight-binding method, we calculate the flux and offset-charge dependence of the spectrum, see Fig.~\ref{fig:compare_tb_ed_flux_qubit}. For the parameters considered, the localization ratios are large compared with unity, indicating that the parameter regimes are amenable to tight binding.
Figs.~\ref{fig:compare_tb_ed_flux_qubit}(a-b) show the spectrum as a function of flux and offset charge, respectively, with improper-tight-binding results overlaying the exact spectrum obtained via charge-basis diagonalization. 
While the spectra from the two different methods are indistinguishable in the upper panels of Figs.~\ref{fig:compare_tb_ed_flux_qubit}(a-b),
we explicitly visualize the residuals for the four lowest eigenergies in the lower panels of Figs.~\ref{fig:compare_tb_ed_flux_qubit}(a-b). For $\Sigma_{\text{max}}=5$, the residuals are all below 1 MHz for flux and offset-charge variation. Further suppression of the absolute error below 1 kHz is possible by increasing the global excitation number cutoff to $\Sigma_{\text{max}}=10$.

\begin{figure}
    \centering
    \includegraphics[width=1.0\columnwidth]{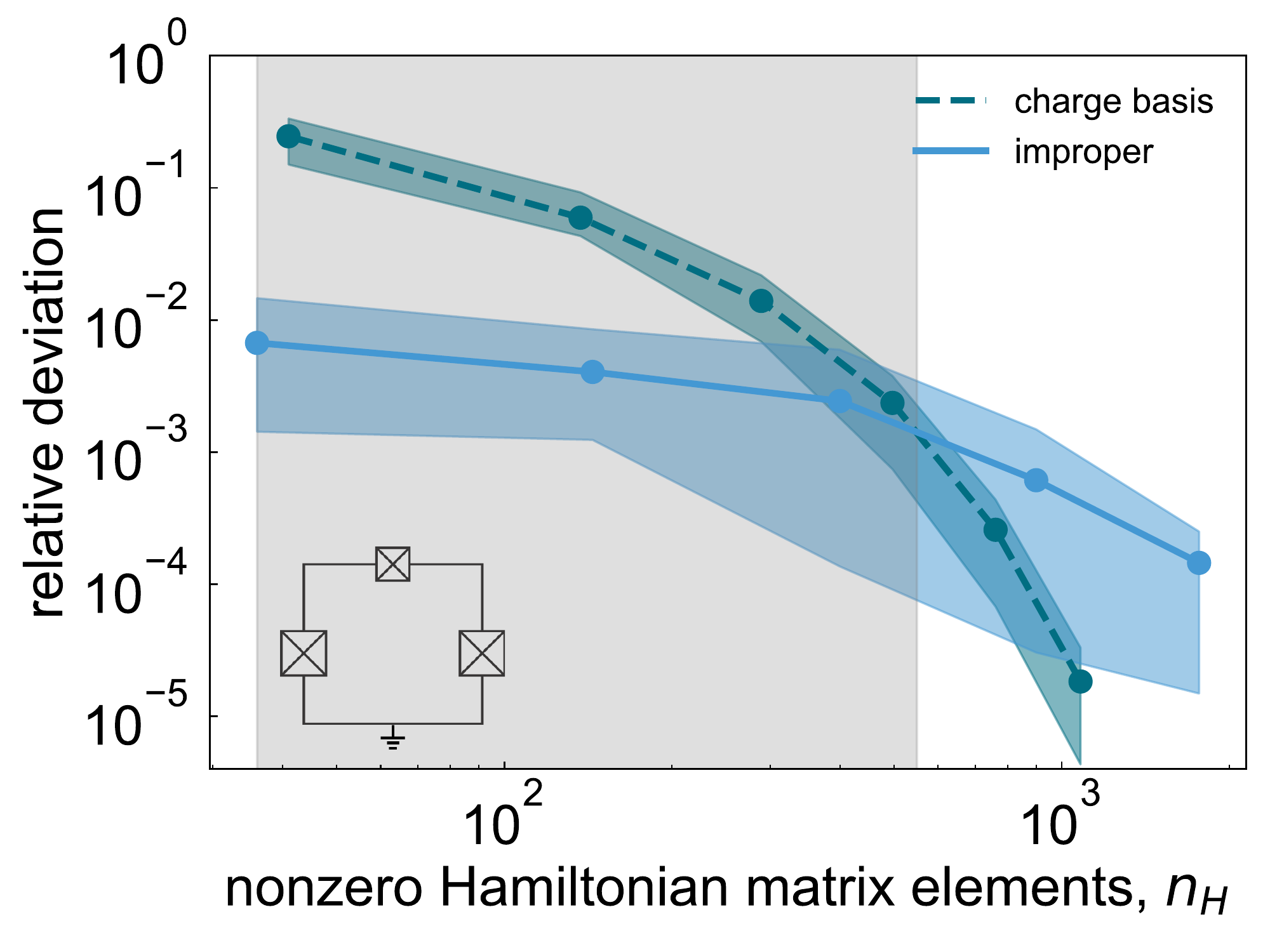}
    \caption{Comparison of convergence to the exact low-energy flux qubit spectrum between improper tight binding (blue, solid) and approximate diagonalization in the charge basis (green, dashed) as a function of $n_{H}$. The colored circles represent the average relative deviation $\eta_{\text{avg}}$, while the colored lines are merely a guide to the eye. The colored shaded regions encompass the range between $\eta_{\text{min}}$ and $\eta_{\text{max}}$. The gray shaded region represents the $n_{H}$ values for which tight binding yields an advantage over charge-basis diagonalization, comparing $\eta_{\text{avg}}$ for a given $n_{H}$. Improper tight binding allows for an accurate estimate of the low-energy eigenspectrum already at $\Sigma_{\text{max}}=1$, yielding $\eta_{\text{avg}} < 7\cdot10^{-3}$ and maximum absolute error of less than $25$ MHz. We choose the parameters of Fig.~\ref{fig:compare_tb_ed_flux_qubit}(a), as well as $\varphi_{\text{ext}}=0.47, n_{g1}=0.2, n_{g2}=0.3$. We can perform the same calculation for the parameters of Fig.~\ref{fig:compare_tb_ed_flux_qubit}(b) and obtain similar results, with tight binding outperforming charge-basis diagonalization for small $n_{H}$. The inset shows a schematic of the flux-qubit circuit.}
    \label{fig:convergence_FQ}
\end{figure}

Even for relatively greedy cutoffs of the global excitation number $\Sigma_{\text{max}}$, the improper-tight-binding method can provide accurate estimates of the eigenspectrum. To compare results obtained using tight-binding methods with results from exact diagonalization, we compute the relative deviation from the exact low-energy spectrum, averaged over the four lowest-energy eigenvalues
\begin{align}
\eta_{\text{avg}} = \frac{1}{4}\sum_{i=0}^{3}\frac{E_{i}-\epsilon_{i}}{\epsilon_{i}}.
\end{align}
Here, $\epsilon_{i}$ is the exact eigenenergy of the state indexed by $i$ and $E_{i}$ is the approximate eigenenergy. We also define the minimum and maximum relative deviations
\begin{align}
\eta_{\substack{\text{min}\\\text{max}}}&=
{\substack{\displaystyle \min\\[1mm]\displaystyle\max\\i=0,\ldots,3}}
\left(\frac{E_{i}-\epsilon_{i}}{\epsilon_{i}}\right). 
\end{align}
To monitor convergence and assess the memory requirements for reaching a desired accuracy, we plot in Fig.~\ref{fig:convergence_FQ} $\eta_{\text{avg}}$ as a function of nonzero Hamiltonian matrix elements ($n_{H}$). We use $n_{H}$ rather than Hilbert-space dimension as a proxy for memory usage to account for the different cases of sparse vs.\ dense matrix numerics encountered for diagonalization in the charge basis vs.\ tight binding. For a cutoff as greedy as $\Sigma_{\text{max}}=1$, we find $\eta_{\text{avg}}<7\cdot10^{-3}$ using improper tight binding. Note that for the flux qubit in the parameter regimes considered here, neither the proper-tight-binding technique nor anharmonicity correction provided any appreciable benefit in terms of convergence to the spectrum over improper tight binding. 

To benchmark convergence of the tight-binding method, we compare against results obtained using {\it truncated} diagonalization in the charge basis. We compute the average relative deviation $\eta_{\text{avg}}$ using energy estimates obtained via a choice of charge-basis cutoff $n_{\text{cut}}$. By increasing $n_{\text{cut}}$, we increase $n_{H}$ and can thereby perform a direct comparison with $\eta_{\text{avg}}$ values obtained via tight binding, see Fig.~\ref{fig:convergence_FQ}. The shaded region of Fig.~\ref{fig:convergence_FQ} indicates where tight binding outperforms approximate diagonalization in the charge basis for a given $n_{H}$. The advantage region for tight binding is for small values of $n_{H}$, indicating that when keeping few basis states, tight-binding states yield a closer approximation to the true low-energy eigenstates than charge-basis states. At larger values of $n_{H}$, charge-basis diagonalization begins to outperform tight binding.

\section{Tight Binding Applied to the Current-Mirror circuit}
\label{sec:currentmirror}

\begin{figure}
    \centering
    \includegraphics[width=1.0\columnwidth]{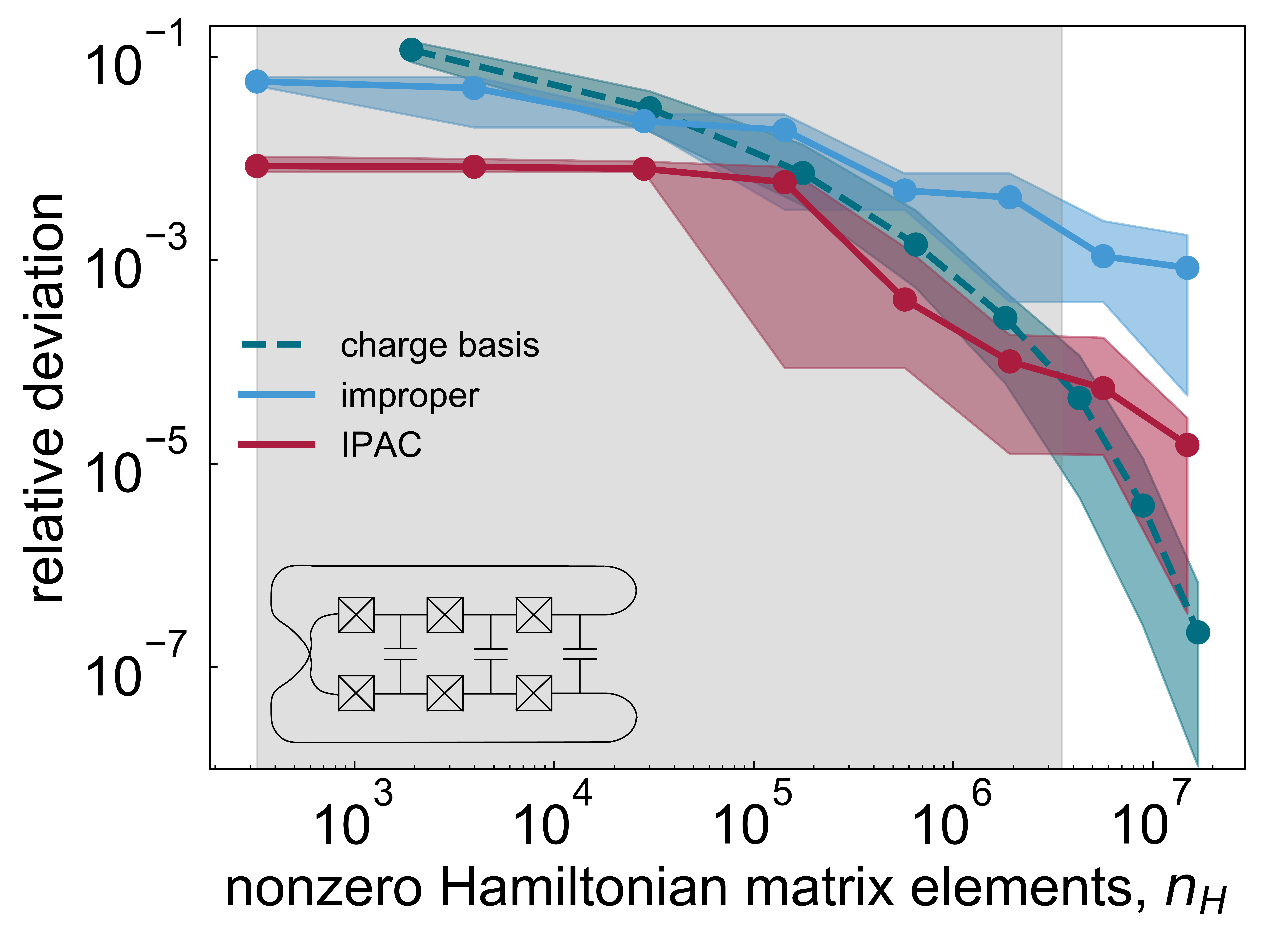}
    \caption{Performance of the tight-binding method as applied to the $N_{B}=3$ current-mirror circuit. Similarly to the case of the flux qubit, we plot $\eta_{\text{avg}}$ for improper tight binding (blue, solid), improper tight binding with anharmonicity correction (red, solid) and approximate diagonalization in the charge basis (green, dashed) as a function of $n_{H}$. Improper tight binding with anharmonicity correction outperforms charge-basis diagonalization across approximately four orders of magnitude in $n_{H}$, as indicated by the shaded region. The sharp cliff in $\eta_{\text{min}}$ for improper tight binding with anharmonicity correction at $n_{H}\approx10^5$ is due to the inclusion of new basis states that contribute to the ground state, yielding $\eta_{\text{min}}\approx10^{-4}$. The inset shows a schematic of the $N_{B}=3$ current mirror circuit. We choose $\varphi_{\text{ext}}=0$ and $n_{gi}=0$, with circuit parameters given in the main text.}
    \label{fig:CM_convergence}
\end{figure}

We expect the tight-binding method to be most useful in the study of larger circuits, where keeping a generous number of basis states is not feasible due to memory requirements.
To demonstrate the tight-binding method on such a larger circuit, we apply it to the  current mirror circuit \cite{Kitaev2006}, described by the Hamiltonian \cite{Weiss2019, DiPaolo2019}
\begin{align}
\label{eq:currentmirrorhamiltonian}
\mathcal{H}_{\text{CM}}=&\sum_{i,j=1}^{2N_{B}-1}(n_{i}-n_{gi})4(\mathsf{E_{C}})_{ij}(n_{j}-n_{gj}) \\ \nonumber 
-&E_{J}\sum_{i=1}^{2N_{B}-1}\cos(\phi_{i}-\varphi_{\text{ext}}/2N_{B}) \\ \nonumber 
-&E_{J}\cos(\Sigma_{i=1}^{2N_{B}-1}\phi_{i}-\varphi_{\text{ext}}/2N_{B})+2N_{B}E_{J},
\end{align}
where $N_{B}$ refers to the number of big capacitors. 
The charging energy matrix $\mathsf{E_{C}}$ involves contributions from individual charging energies $E_{C_{B}}, E_{C_{J}}, E_{C_{g}}$ due to the big-shunt, junction and ground capacitances respectively \cite{Weiss2019}. An example circuit with $N_{B}=3$ without the capacitors to ground is shown in the inset of Fig.~\ref{fig:CM_convergence}. The number of degrees of freedom of the circuit is given by $2N_{B}-1$. The interest in this circuit originates from Kitaev's prediction that quantum information should be protected relaxation and dephasing in the current mirror \cite{Kitaev2006}. For a representative choice of parameters, one can identify $N_{B}\approx12$ as the ideal value of $N_{B}$ \cite{Weiss2019}. Circuit sizes with such large values of $N_{B}$ exceed our capabilities for finding eigenstates and eigenenergies via diagonalization in the charge basis; the maximum value of $N_{B}$ where we can achieve spectral convergence is $N_{B}=3$.\footnote{Our calculations were performed on an Intel Xeon CPU E5-1650 24 core processor with 128 GB RAM.} We show below that the tight-binding method is an advantageous alternative for simulating the current-mirror circuit at larger values of $N_{B}$.

Implementation of the tight-binding method for the current-mirror circuit proceeds in a manner analogous to the case of the flux qubit, as neither circuit contains inductors. We choose a set of protected circuit parameters 
given by $E_{C_{B}}/h=0.2$ GHz, $E_{C_{J}}/h=35$ GHz, $E_{C_{g}}/h=45$ GHz, $E_{J}/h=10$ GHz, $\varphi_{\text{ext}}=0$ and $n_{gi}=0$.
To establish qualitatively that the current-mirror circuit with these parameters is amenable to a tight-binding treatment, we fix a value of $N_{B}$ and verify that the localization ratios are all of order unity or larger. We observe that the localization ratios generally {\it increase} with $N_{B}$, indicating that the tight-binding method should become increasingly accurate with larger $N_{B}$. For an independent quantitative assessment of the validity of tight binding, we will compare spectra obtained with tight-binding methods with exact results. For this purpose we first apply the tight-binding method to the $N_{B}=3$ current-mirror circuit. 

We can obtain excellent agreement between spectra obtained via tight binding and exact results, with average relative deviations $\eta_{\text{avg}}$ below $2\cdot10^{-5}$, see Fig.~\ref{fig:CM_convergence}. The best agreement is for the energy of the ground state, for which we obtain agreement to within 16 kHz. For the first- and second-excited states these results correspond to sub-MHz agreement, while for the third-excited state agreement is on the order of a MHz. The use of the anharmonicity correction yields a substantial benefit that is critical for achieving this level of accuracy. We find that the proper-tight-binding method yields nearly identical results to those produced by improper tight binding, and therefore those results are not shown in Fig.~\ref{fig:CM_convergence}. Our highest accuracy approximations are obtained with $\Sigma_{\text{max}}=8$, beyond which we encounter numerical instabilities. We emphasize that one can actually obtain a reasonable approximation to the spectrum based on moderate values of $\Sigma_{\text{max}}$, as shown in Fig.~\ref{fig:CM_convergence}. For example with $\Sigma_{\text{max}}=1,2$, improper tight binding with anharmonicity correction yields $\eta_{\text{avg}} \approx 8\cdot10^{-3}$, corresponding to absolute errors of about $300$ MHz.

We can contrast these results with those obtained using truncated diagonalization in the charge basis. Using the same metric for memory efficiency previously applied to the flux-qubit example, we find that tight binding is advantageous over a wide range of $n_{H}$ values, 
see Fig.~\ref{fig:CM_convergence}. Specifically, to achieve $\eta_{\text{avg}}\approx8\cdot10^{-3}$, truncated diagonalization in the charge basis requires about three more orders of magnitude in memory resources as compared to tight binding. The advantage region for tight binding extends over approximately four orders of magnitude $10^2 \lesssim n_{H} \lesssim 10^6$, as shown in the shaded area of Fig.~\ref{fig:CM_convergence}. 

To extend toward the regime of ideal $N_{B}$, we apply the tight-binding method to obtain the spectrum of the $N_{B}=5$ current-mirror circuit, which has 9 degrees of freedom. We compute the ground-state energy $E_{0}$ and first-excited-state energy $E_{1}$ using the four tight-binding techniques [improper, proper, (IPAC) improper with anharmonicity correction and (PAC) proper with anharmonicity correction of the $m=0$ minimum], see Tab.~\ref{table:CM_N5_evals}. By the variational principle, our computed eigenenergies are upper bounds to the true eigenenergies \cite{Macdonald1933, Bishop1989, Lowdin1967}. Therefore, lower eigenenergy values always imply higher accuracy. The proper, IPAC and PAC tight-binding schemes all perform similarly but collectively outperform the improper scheme.
The lowest eigenenergies are obtained using IPAC, with bounds $\epsilon_{0}\leq81.6472$ GHz and $\epsilon_{1}\leq82.7224$ GHz. The largest cutoff we can handle is $\Sigma_{\text{max}}=5$, beyond which we encounter memory issues.
We observe that for this circuit, ansatz states localized in minima aside from the $m=0$ minimum contribute to the low-energy spectrum, and moreover the curvatures of those minima differ from those of the $m=0$ minimum. Otherwise, there would be no difference between the eigenenergies computed with improper and proper tight binding.
Note that schemes IPAC, PAC allow for rough estimates of the eigenspectrum with a greedy cutoff $\Sigma_{\text{max}}=2$, with calculated $E_{0}, E_{1}$ less than $200$ MHz greater than the lowest obtained respective values.

\begin{figure}
    \centering
    \includegraphics[width=1.0\columnwidth]{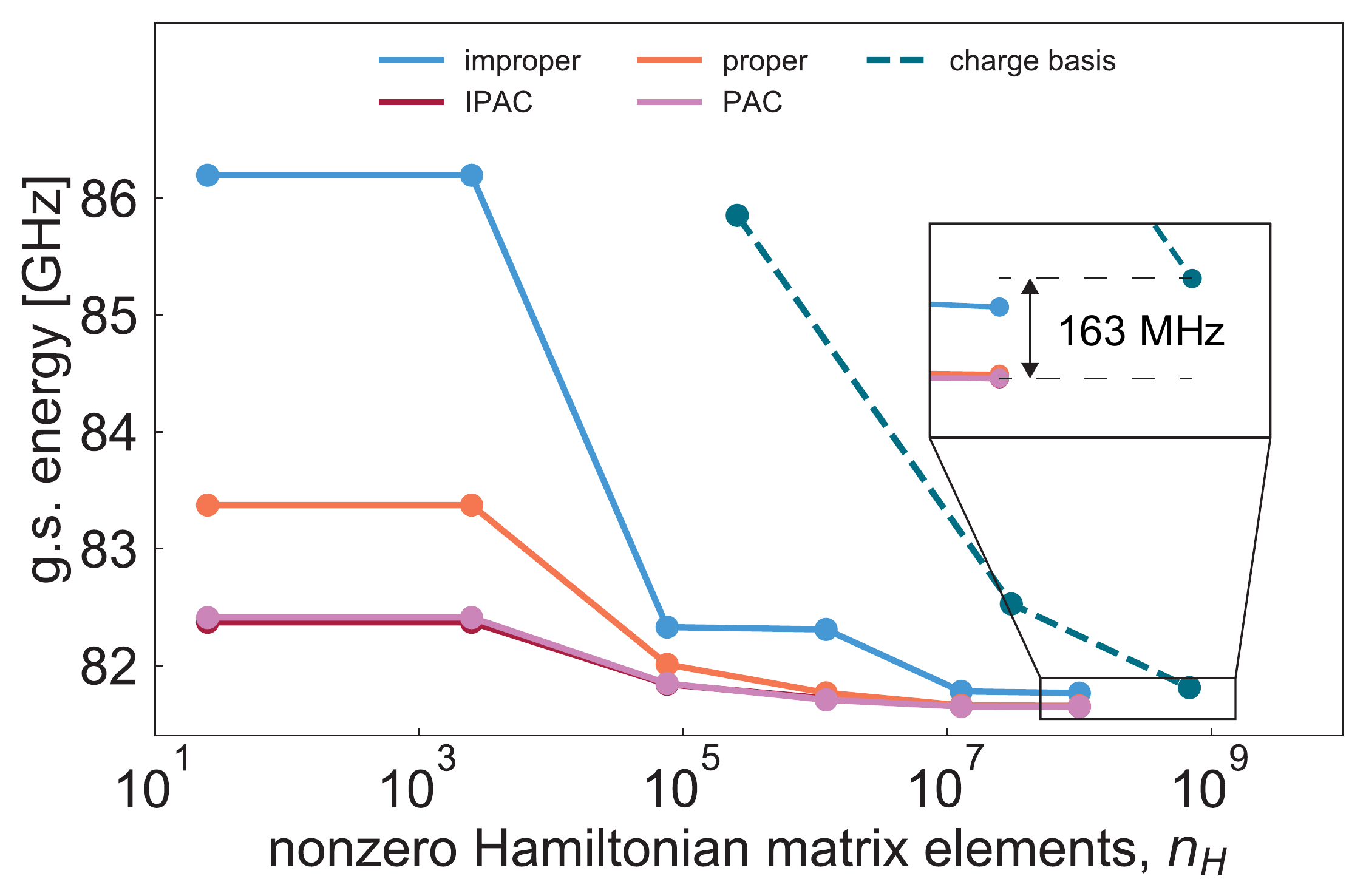}
    \caption{Comparison of computed ground-state energies for the $N_{B}=5$ current-mirror circuit. Individual curves correspond to results obtained with tight-binding schemes improper (blue, solid), improper with anharmonicity correction (red, solid), proper (orange, solid) and proper with anharmonicity correction of the global minimum (pink, solid), as well as approximate diagonalization in the charge basis (green, dashed). Tight-binding techniques consistently yield lower and hence more accurate eigenenergies as compared to charge basis diagonalization, with a difference of 163 MHz between the best results obtained with tight binding and approximate diagonalization in the charge basis, see the inset. We used the same current-mirror circuit parameters here as in Fig.~\ref{fig:CM_convergence}. }
    \label{fig:CM_5}
\end{figure}

We next compare tight-binding results with those from approximate diagonalization using the truncated charge basis. For $N_{B}=5$ the maximum possible charge cutoff we can handle is $n_{\text{cut}}=3$, corresponding to a Hilbert-space dimension of $(2\cdot n_{\text{cut}}+1)^{9}=7^9\approx4\cdot10^7$. The best estimates for $E_{0}$ and $E_{1}$ obtained using approximate diagonalization in the charge basis are in fact {\it higher} and therefore less accurate than the lowest obtained values using tight-binding methods, see Tab.~\ref{table:CM_N5_evals}. Moreover, the tight-binding methods consistently yield lower eigenenergy approximations across all $n_{H}$ values. Fig.~\ref{fig:CM_5} illustrates this point for the ground-state energy $E_{0}$, and similar results hold for the first-excited-state energy $E_{1}$. We thus find that the tight-binding method is more memory efficient than charge-basis diagonalization for the $N_{B}=5$ current-mirror circuit. More broadly, this may indicate that the tight-binding method can serve as an interesting and useful method in the context of large circuits. 

\begin{table*}
\includegraphics[width=2\columnwidth]{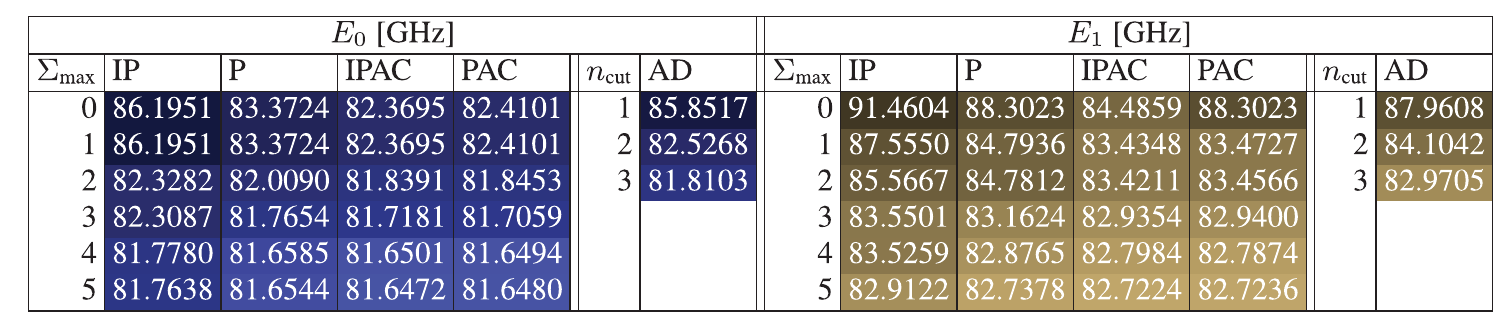}
\caption{\label{table:CM_N5_evals}Eigenenergies for the ground state and first-excited state of the $N_{B}=5$ current-mirror circuit. Energies were computed using tight-binding schemes (IP) improper, (P) proper, (IPAC) improper with anharmonicity correction, (PAC) proper with anharmonicity correction, as well as (AD) approximate diagonalization in the charge basis. The energies are color coded from least accurate (darkest) to most accurate (lightest). The three tight-binding flavors proper, IPAC and PAC all perform similarly and outperform improper. The most accurate results for $E_{0}$ and $E_{1}$ were obtained with tight binding rather than with approximate diagonalization in the charge basis (circuit parameters used are the same as in Fig.~\ref{fig:CM_convergence}).}
\end{table*}

\section{Conclusion}
\label{sec:conclusion}

We have generalized the well-known method of tight binding for the purpose of efficiently and accurately obtaining the low-energy spectra of superconducting circuits. 
Construction of the Hamiltonian proceeds by using ansatz Bloch states that localize in minima of the potential. The method can handle many degrees of freedom, multiple inequivalent minima, and periodic or extended potentials. In terms of these states, the Schr\"odinger equation turns into a generalized eigenvalue problem. Solving it yields a spectrum that provides upper bounds to the true eigenenergies. To establish the accuracy of the tight-binding method we apply it to the flux qubit and achieve agreement with exact results at the kHz level. 

Because the method is expected to be of use for larger circuits, we apply it to the $N_{B}=3, 5$ current-mirror circuits, which have 5 and 9 degrees of freedom, respectively. We find excellent agreement with exact results in the case of the $N_{B}=3$ circuit. Moreover, across multiple orders of magnitude in memory usage (as quantified by $n_{H}$), eigenenergies computed using tight binding are found to be more accurate than those calculated using the charge basis. For the $N_{B}=5$ circuit, the tight-binding method allows for the extraction of eigenenergies that are lower than any obtainable using the charge basis, given our computational resources. This work supplements recent research also focused on the efficient simulation of large superconducting circuits \cite{Kerman2020, ding2020}.

To extend and improve the tight-binding method beyond what is described here, we envision developing an improved state-optimization procedure beyond optimizing the harmonic lengths of the ansatz ground state, as well as devising a hybrid method including both tight binding and charge-basis diagonalization to accommodate circuits with both localized and delocalized degrees of freedom.

\section{Acknowledgements}

The authors thank Z. Huang and X. You for insightful discussions.
D.~K.~W. was supported with a QuaCGR Fellowship by the Army Research Office (ARO) . This research was funded by the ARO under Contract No. W911NF-17-C-0024.
 
\appendix
\section{Normal ordering in the presence of squeezing}
\label{appendix:squeezing}

As discussed in the main text, we must normal order the operator product
\begin{align}
\label{eq:squeezing_opprod}
\mathcal{S}_{m'}^{\dagger}\mathcal{T}_{\vec{\theta}_{m'}}^{\dagger}\mathcal{O}\,\mathcal{T}_{\vec{\theta}_{m}+2\pi\vec{j}}\,\mathcal{S}_{m}
\end{align}
prior to numerical evaluation. Normal ordering of the squeezing operator $\mathcal{S}_{m}$ [Eq.~\eqref{eq:squeezingop}] proceeds by first placing $\mathcal{S}_{m}$ in so-called disentangled form
\cite{Balian1969, Wang1994}
\begin{align}
\label{eq:squeezingop_disentangled}
\mathcal{S}_{m}=&\exp(-\tfrac{1}{2}\Tr{Y})
\exp(-\tfrac{1}{2}a_{\mu}^{\dagger}X_{\mu\nu} a_{\nu}^{\dagger}) \\ \nonumber 
\times&\exp(-a_{\mu}^{\dagger}Y_{\mu\nu} a_{\nu})
\exp(\tfrac{1}{2}a_{\mu}Z_{\mu\nu}a_{\nu}).
\end{align}
where $X=u^{-1}v, Y=\ln{u}, Z=vu^{-1}$, where we omit the $m$ dependence of these quantities for notational simplicity \cite{Javanainen}.
The inner term of Eq.~\eqref{eq:squeezingop_disentangled} with $Y$ is not yet normal ordered. This can be rectified via the formula \cite{Mehta}
\begin{align}
\label{eq:norm_adag_a}
\exp(a_{\mu}^{\dagger}Y_{\mu\nu}a_{\nu})=\;:\exp(a_{\mu}^{\dagger}(e^Y-\openone)_{\mu\nu}a_{\nu}):,
\end{align}
where $:\,:$ is known as the normal-ordering symbol. Creation and annihilation operators inside the normal-ordering symbol can be commuted without making use of the commutation relations. A trivial example of the use of this superoperator is $:aa^{\dagger}:\;=a^{\dagger}a$. 

To commute exponentials and $\mathcal{V}$ operators appearing for example in Eq.~\eqref{eq:squeezingop_disentangled} and Eq.~\eqref{eq:V} we make use of the following normal-ordering formulae \cite{Ma1990, Hongyi}
\begin{widetext}
\begin{align}
\label{eq:witschel_1_multi}
\exp(a_{\mu}Z_{\mu\nu}a_{\nu})\exp(a_{\mu}^{\dagger}X_{\mu\nu}a_{\nu}^{\dagger})&=
\frac{1}{\sqrt{\det(\openone-4ZX)}} \exp(a_{\mu}^{\dagger}[(\openone-4XZ)^{-1}X]_{\mu\nu}a_{\nu}^{\dagger})\\ \nonumber 
&\times\exp(a_{\mu}^{\dagger}[\ln(\openone-4XZ)^{-1}]_{\mu\nu}a_{\nu}) \exp(a_{\mu}[(\openone-4ZX)^{-1}Z]_{\mu\nu}a_{\nu}), 
\\ 
\label{eq:danny_adag_a}
\exp(a_{\mu}^{\dagger}Y_{\mu\nu}a_{\nu})\exp(a_{\mu}^{\dagger}Y'_{\mu\nu}a_{\nu}) &= \exp(a_{\mu}^{\dagger}\ln(e^Ye^{Y'})_{\mu\nu}a_{\nu}), \\
\label{eq:danny_1_multi}
\exp(a_{\mu}Z_{\mu\nu}a_{\nu})\exp(\lambda_{\mu}a_{\mu}^{\dagger}) &= \exp(\lambda_{\mu}Z_{\mu\nu}\lambda_{\nu}) 
\exp(\lambda_{\mu}a_{\mu}^{\dagger})
\exp(a_{\mu}Z_{\mu\nu}a_{\nu})\exp(\lambda_{\mu}(Z_{\mu\nu}+Z_{\mu\nu}^{t})a_{\nu}), \\
\label{eq:wilcox10_20_1_multi}
\exp(a_{\mu}^{\dagger}Y_{\mu\nu}a_{\nu})\exp(\lambda_{\mu}a_{\mu}^{\dagger}) 
&=
\exp(a_{\mu}^{\dagger}(e^{Y})_{\mu\nu}\lambda_{\nu}) \exp(a_{\mu}^{\dagger}Y_{\mu\nu}a_{\nu}), \\
\label{eq:wilcox10_20_2_multi} 
\exp(a_{\mu}^{\dagger}Y_{\mu\nu}a_{\nu})\exp(a_{\mu}^{\dagger}X_{\mu\nu}a_{\nu}^{\dagger})&=
\exp(a_{\mu}^{\dagger}(e^{Y})_{\mu\nu}X_{\nu\sigma}(e^{Y})^{t}_{\sigma\tau}a_{\tau}^{\dagger}) \exp(a_{\mu}^{\dagger}Y_{\mu\nu}a_{\nu}).
\end{align}
Here, $X, Y, Y', Z$ and $\vec{\lambda}$ are arbitrary, except for the requirement of $\openone-4XZ$ and $\openone-4ZX$ to be nonsingular in Eq.~\eqref{eq:witschel_1_multi}.
We note that it is relatively straightforward to obtain Eqs.~\eqref{eq:danny_1_multi}-\eqref{eq:wilcox10_20_2_multi} from standard applications of the BCH formula \cite{Wilcox}. Obtaining Eqs.~\eqref{eq:witschel_1_multi}-\eqref{eq:danny_adag_a} is slightly more difficult, and requires using either Lie algebra techniques \cite{Ma1990, CoherentZhang1990} or the so-called IWOP procedure \cite{Hongyi}.

An instance of Eq.~\eqref{eq:squeezing_opprod} relevant for computing wavefunction overlaps is $\mathcal{S}_{m'}^{\dagger}\exp (\vec{\lambda}^{t}\vec{a^{\dagger}})\exp(-\vec{\lambda}^{t}\vec{a})\mathcal{S}_{m}$, identifying
\begin{align}
\vec{\lambda} &= \frac{1}{\sqrt{2}}(\vec{\theta}_{m}-\vec{\theta}_{m'}+2\pi\vec{j})^{t}\Xi^{-t},
\end{align}
and neglecting the overall multiplicative factor [c.f. Eq.~\eqref{eq:BCH}]. To simplify notation, we have suppressed the dependence of $\vec{\lambda}$ on $m, m'$ and $\vec{j}$. We will continue to likewise suppress the $m$ dependence of the various matrices and distinguish between $X_{m}$ and $X_{m'}$ by using the notation $X$ and $X'$, etc. Applying each of the relations Eq.~\eqref{eq:witschel_1_multi}-\eqref{eq:wilcox10_20_2_multi} in a few steps of algebra leads to the normal-ordered result
\begin{align}
\mathcal{S}_{m'}^{\dagger}\exp(\vec{\lambda}^{t}{\vec{a^{\dagger}}})\exp(-\vec{\lambda}^{t}\vec{a})\mathcal{S}_{m} = &\frac{\exp(-\frac{1}{2}[\vec{\lambda}^{t}\{X+(\openone+X)\overline{P^{t}X'}(\openone+X)\}\vec{\lambda}+\Tr{Y'^{\dagger}}+\Tr{Y}])}{\sqrt{\det(\openone-X' X)}}
\\ \nonumber 
\times  
&\exp(-\tfrac{1}{2}\vec{a^{\dagger}}^{t}[\{e^{-Y'}\}^{\dagger}PX\{e^{-Y'}\}^{*}-Z']\vec{a^{\dagger}})\exp(\vec{\lambda}^{t}[\openone+X]P^{t}[e^{-Y'}]^{*}\vec{a^{\dagger}})
\\ \nonumber 
\times 
&:\exp(\vec{a^{\dagger}}^{t}[e^{-Y'^{\dagger}}P e^{-Y}-\openone]\vec{a}):
\\ \nonumber 
\times
&\exp(-\vec{\lambda}^{t}[\openone+\{\openone+X\}\overline{P^{t}X'}]e^{-Y}\vec{a})\exp(\tfrac{1}{2}\vec{a}^{t}[Z-\{e^{-Y}\}^{t}P^{t}X' e^{-Y}] \vec{a}),
\end{align},
\end{widetext}
where $P=(\openone-XX')^{-1}$, $\overline{P^{t}X'} = \frac{1}{2}(P^{t}X'+X'P)$ and the matrices $X, X'$, etc. can be taken to be symmetric. Similar expressions can be obtained when the operator $\mathcal{O}$ is an explicit function of the ladder operators $\vec{a}, \vec{a^{\dagger}}$. 

\bibliography{bib}

\end{document}